# Relativistic electrons from sparks in the laboratory

N. Østgaard[1], B. E. Carlson[1,2], R. S. Nisi[1], T. Gjesteland[1], Ø. Grøndahl[1], A. Skeltved[1], N. G. Lehtinen[1], A. Mezentsev[1], M. Marisaldi[1,3], and P. Kochkin[1,4]

[1]Birkeland Centre for Space Science, Department of Physics and Technology, University of Bergen, Bergen, Norway, [2]Department of Physics, Carthage College, Kenosha, Wisconsin, USA, [3]INAF-IASF, National Institute for Astrophysics, Bologna, Italy, [4]Department of Electric Engineering, Technische Iniversiteit Eindhoven, Eindhoven, Netherlands

**Abstract** Discharge experiments were carried out at the Eindhoven University of Technology in 2013. The experimental setup was designed to search for electrons produced in meter-scale sparks using a 1 MV Marx generator. Negative voltage was applied to the high voltage (HV) electrode. Five thin (1 mm) plastic detectors (5 cm$^2$ each) were distributed in various configurations close to the spark gap. Earlier studies have shown (for HV negative) that X-rays are produced when a cloud of streamers is developed 30–60 cm from the negative electrode. This indicates that the electrons producing the X-rays are also accelerated at this location, that could be in the strong electric field from counterstreamers of opposite polarity. Comparing our measurements with modeling results, we find that ~300 keV electrons produced about 30–60 cm from the negative electrode are the most likely source of our measurements. A statistical analysis of expected detection of photon bursts by these fiber detectors indicates that only 20%–45% of the detected bursts could be from soft (~10 keV) photons, which further supports that the majority of detected bursts are produced by relativistic electrons.

## 1. Introduction

Since the discovery of terrestrial gamma ray flashes (TGFs) from lightning strokes by The Burst And Transient Source Experiment (BATSE) [*Fishman et al.*, 1994], there has been a renewed interest in studying X-rays from discharges in laboratory [*Dwyer et al.*, 2005, 2008; *Nguyen et al.*, 2008; *Kochkin et al.*, 2012, 2014, 2015]. The advantage of studying discharges in a laboratory is that many more details can be measured, even if the spatial and temporal scale is smaller. With the particle detectors and fast cameras that are now available, even the time scales of nanoseconds can be studied [*Kochkin et al.*, 2012, 2014]. However, there are important differences between natural lightning and discharges in the laboratory.

Studies from the last 10 years strongly indicate that TGFs are produced during the initial phase of intracloud positive (IC+) lightning bringing negative charges upward [*Cummer et al.*, 2005; *Stanley et al.*, 2006; *Shao et al.*, 2010; *Lu et al.*, 2010; *Cummer et al.*, 2011; *Østgaard et al.*, 2013; *Cummer et al.*, 2015]. Typical potential differences in IC+ lightning range from 10 MV to 100 MV [*Rakov and Uman*, 2003]. The theories for explaining the production of TGFs consequently consider potentials of tens of megavolt (MV). Important building blocks of these theories are the relativistic runaway electron process [*Wilson*, 1925], which requires electrons of several tens of keV for typical large-scale electric fields to become runaway. Furthermore, the relativistic runaway electron avalanche (RREA) suggested by *Gurevich et al.* [1992] requires tens of MV potential difference in order to be consistent with the high energies of observed $\gamma$ rays. An avalanche length can be understood as the length over which a sufficient potential drop exists to give a multiplication of a factor of $e$. The potential drop over an avalanche length ranges from 7 MV to 10 MV, depending on the electric field strength [*Coleman and Dwyer*, 2006; *Skeltved et al.*, 2014]. To produce the number of electrons and consequently number of photons in a TGF, one would need 10–20 avalanche lengths, given a seed population of $10^{10}$ electrons. This corresponds to a total potential of 70–140 MV. If the number of seed electrons is larger, fewer avalanche lengths are needed. To obtain the typical RREA spectrum, which reaches a self-similar state with a characteristic exponential falloff, one would need 5–6 avalanche lengths. Also, the feedback mechanism, where positrons and scattered gamma photons can produce new avalanches, suggested by *Dwyer* [2012], would need potentials >100 MV to be efficient [*Dwyer*, 2012; *Skeltved et al.*, 2014].

In the electric discharge experiments performed in the laboratory the potentials are typically about 1 MV. This means that neither RREA nor feedback can be produced in laboratory experiments. What we expect to observe in the laboratory is the formation of streamers and leaders, and since the electric fields and potential drops in the tip of streamers are large enough to accelerate electrons into the runaway regime [*Wilson*, 1925; *Moss et al.*, 2006; *Li et al.*, 2009; *Chanrion and Neubert*, 2010], we should be able to see electrons with hundreds of keV.

Earlier studies of laboratory sparks have indeed shown that X-rays are produced and estimates of photon energies range from 30 keV–150 keV [*Dwyer et al.*, 2005] to 200 keV [*Kochkin et al.*, 2015] to <230 keV [*Dwyer et al.*, 2008]. While *Kochkin et al.* [2015] studied discharges with negative voltage applied to the HV electrode, *Dwyer et al.* [2005, 2008] applied both polarities. These energy estimates are fairly uncertain, because the X-ray detectors used in these experiments did not have the temporal resolution to separate single photons and most probably indicate the average energy of the photon bursts (with duration of tens of nanoseconds) rather than the energy of single photons. A statistical analysis that attempts to correct for pile-up effects gives an average energy of 86 keV [*Carlson et al.*, 2015].

Using fast cameras to image the streamer formation during negative discharges and sensitive X-rays detector about 2 m from the sparks *Kochkin et al.* [2014] were able to determine that the X-rays were produced during the fourth burst of streamer formation. To be even more precise, they were produced during the rise time of the fourth current pulse at the HV electrode, which is the first half of the pulse [see *Kochkin et al.*, 2015, Figure 2]. At this stage the streamer zone had reached a distance of 30–60 cm away from the HV electrode [*Kochkin et al.*, 2014, Figure 6]. By using two cameras with consecutive exposure time of 50 ns, they could also see streamers moving toward the HV electrode in this region coinciding in time with the X-ray observation. Based on these observations they suggested that the strong local electric field between streamers encountering each other could be the acceleration region for the electrons that eventually produce the X-rays. We will call these counterstreamers throughout this paper. It should be noticed that both the source location and production time of the X-rays are quite different when positive voltage is applied to the HV electrode [see *Kochkin et al.*, 2012].

Inspired by the results by *Dwyer et al.* [2008], it had already been suggested some years earlier that counterstreamers could be the source region for the X-rays [*Cooray et al.*, 2009]. They presented a simple model of field between counterstreamers developing in a potential of 850–980 kV between the electrodes. They estimated that the local electric fields between these counterstreamers, as well as between the two streamer zones, were able to accelerate electrons up to 360 keV. Their estimates varied depending on how large the gaps between the single streamers (5 mm to 5 cm) and the length of the streamer zones (30 cm to 70 cm) were. These results have been disputed by *Ihaddadene and Celestin* [2015], who claim that the electric field in the gap of counterstreamers will only produce a small number of electrons.

The images from *Kochkin et al.* [2014] indicate that the counterstreamers developed and the X-rays were produced already when only a negative streamer region was formed. The streamer region from the positive (grounded) electrode developed a few hundreds of nanoseconds later. Nevertheless, the electrons energies estimated by *Cooray et al.* [2009] will be relevant for the present study, as will be shown.

Our experiments were performed in the laboratory of Technical University of Eindhoven and the overall experimental setup was similar as used by *Kochkin et al.* [2014]. Two electrodes were placed vertically 107 cm apart, and a Marx generator produced a 1 MV potential between the electrodes with a rise time of about 1 μs. Instead of using sensitive X-ray detectors far from the sparks (as in *Kochkin et al.* [2012, 2014] and *Carlson et al.* [2015]), we designed detectors that should be more sensitive to electrons than X-rays and intentionally placed them as close to the spark as possible. The goal was to detect the electrons that were produced in the streamer zone.

The questions we want to address are the following:

1. Are we able to distinguish electrons from X-rays?
2. Where are the electrons accelerated?
3. What are the energies of the electrons?

## 2. Experimental Setup

As mentioned, we designed detectors that should be much more sensitive to electrons than to photons. To achieve this, we used scintillating plastic fibers, 1 mm diameter and 10 cm long. Five of these were bundled together with a thin tape (0.15 mm), which covers them with a single layer on both sides, to create a detector area of 5 cm$^2$ (5 × 10 cm × 0.1 cm) but only 0.1 cm thick. These bundled scintillating plastic fibers comprise what will be referred to as one detector. This detector was then glued onto a 3 m optical fiber which was connected on the other end to a photomultiplier tube (PMT) which was shielded from electromagnetic interference. The connection between the plastic fiber and optical fiber was supported by a thick tape partially covering the end of the detector, so the detector area was probably closer to 4.5 cm$^2$ than 5 cm$^2$. We made five such detectors, each glued to its own optical fiber and connected to its own PMT that was placed in an electromagnetic shielded cabinet. These five detectors will be referred to as UB1, UB2, UB3, UB4, and H1.

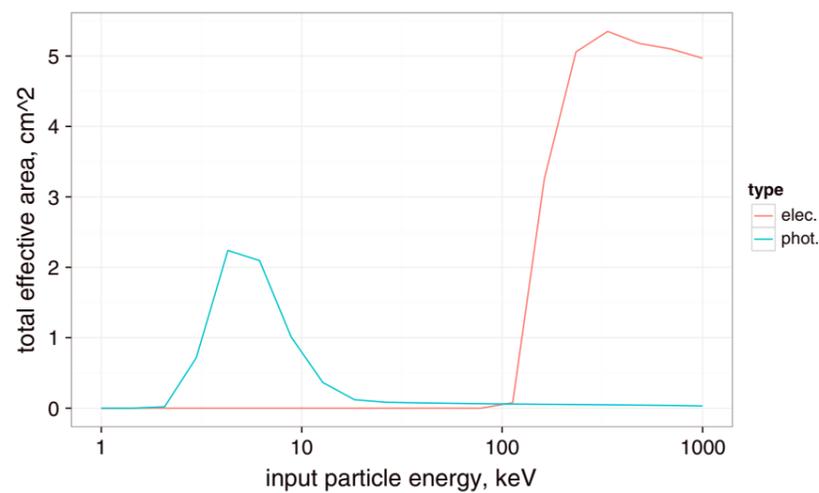

**Figure 1.** A GEANT4 modeling of the detector sensitivity for electrons and photons. The detectors were covered by a tape of 0.15 mm in this simulation.

A result of the GEometry ANd Tracking 4 (GEANT4) modeling of the detector sensitivity is shown in Figure 1. The effective areas are calculated for deposition of more than 1 keV in the scintillator. As photon energy increases, the effective area first increases as the cross section for interaction in the tape decreases, allowing photons to enter the scintillator, but then decreases as interactions in the scintillator become less likely. Similarly, the electron effective area is negligible below 100 keV because such electrons are unable to penetrate the tape. Above 200 keV, the effective area is approximately constant since such particles incident on the detector will always deposit some energy above the detector threshold. One can see that with a 0.15 mm tape, the detectors are sensitive to photons between 3 keV and 15 keV, with an effective area that peaks at about 2 cm$^2$ but average in the interval of about 1 cm$^2$. For electrons of ≥ 200 keV the effective area is about 5 cm$^2$ (or 4.5 cm$^2$ as already mentioned). The detectors are not sensitive to electrons ≤100 keV.

The overall experimental setup is similar to what is shown in Figure 1 in *Kochkin et al.* [2014]. The electrodes were 107 cm from each other vertically, and negative HV was applied to the upper electrode. However, our five detectors were arranged in three different configurations as shown in Figure 2. For each of the configurations we performed 100 sparks. The configurations were azimuthal (red), radial (green), and polar (blue). Table 1 lists the locations of the five detectors relative to the ground electrode for each of the configurations. Each coordinate refers to the center of each detector.

### 3. Data and Signal Recognition

Figure 3 shows typical signals from one discharge. Figures 3 (first panel) to 3 (eighth panel) show the currents at the HV electrode ($I_{HV}$), the ground electrode ($I_{gnd}$) and the Marx generator voltage (V) together with the signals in all five detectors (UB1–UB4 and H1). The signals measured by our detectors typically occur between 0.5 μs and 1 μs, while the spark goes off about 1.4–1.5 μs, when the currents at both electrodes saturate and the HV drops. Time resolution of data is 0.1 ns for UB1–UB4 and 0.2 ns for H1. In all the experiments a negative high voltage of about 1 MV was applied. For this specific spark we detected two distinct bursts (peaks) and all detectors saturated. One can see that UB1–UB3 and H1 have almost similar sensitivity, while UB4 sensitivity is a factor of ∼10 lower. The gain of the detectors was held fixed for all the sparks performed. How we handle the saturated peaks will be discussed in section 4. Throughout this paper we will use the following terminology: (1) spark: each experiment, i.e., a single high-voltage discharge; (2) peak: an above background peak in the detector data; (3) burst: a momentary flux of high-energy particles, which may be detected (or not detected) in the form of simultaneous peaks in the detectors; there can be zero, one, or more bursts produced in a spark;

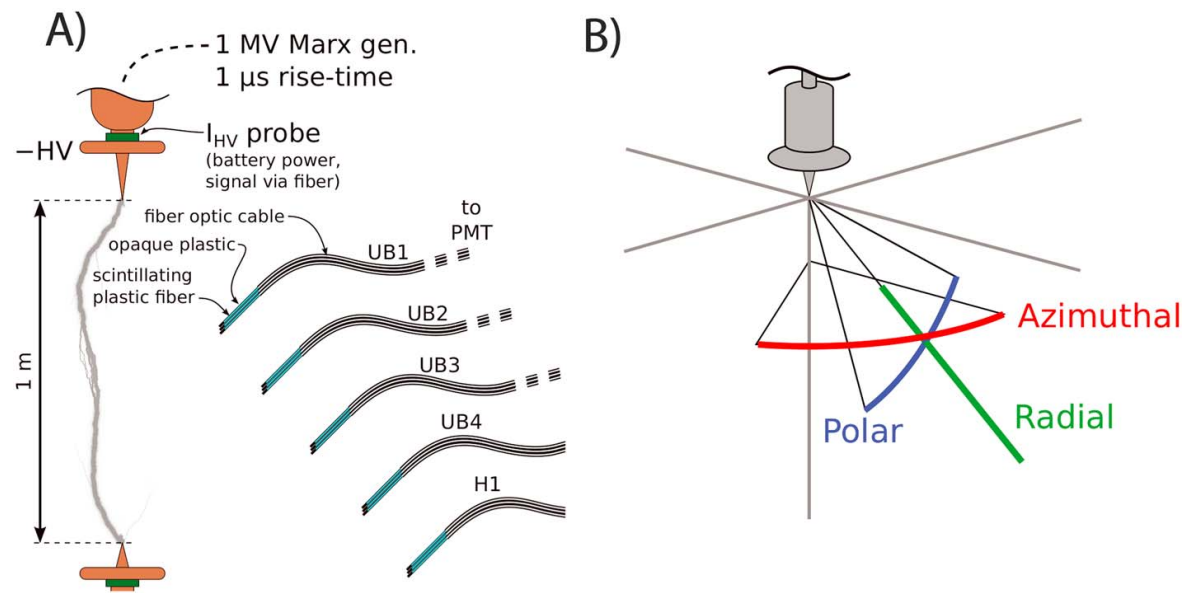

**Figure 2.** (a) The five detectors in a radial configuration. (b) The three arrangements of detectors. Red, azimuthal; green, radial; and blue, polar.

each burst (peak) can be one or more particles which can come from one or more source locations within the 30 ns window; (4) signal: one or more bursts detected in a spark.

To identify signals above the noise level, we proceed as follows:

1. The data from 0 μs to 0.2 μs are used to estimate the standard deviation ($\sigma$) of the background.
2. Due to occasional optical leakage from the bright spark (after 1.4 μs) into the detectors only data from 0 μs to 1.2 μs are used to identify peaks.
3. Data are smoothed, with 14 ns boxcar average.
4. A burst is identified when the absolute value of the smoothed data exceeds $10\sigma$.
5. For each burst we identify the following: (a) start time, that is, when signal reaches more than $4\sigma$; (b) peak time; (c) peak value (of the nonsmoothed data); and (d) the duration of saturated peaks.

Due to ringing in the detectors it was not possible to determine the stop time precisely. For the correction of saturated peaks, which will be described in section 4, we therefore decided to assume a Gaussian (symmetric) shape of the signal, where stop time is peak time + (peak time minus start time).

## 4. Saturation Correction and Calibration

Unfortunately, we did not have the absolute calibration of the detectors, but for the analysis presented in this paper it was sufficient to intercalibrate the detectors. To perform an intercalibration of the five detectors (UB1–UB4 and H1), they were bundled together at a position $X = 63$ cm, $Y = 0$ cm, and $Z = 55$ cm. In Figure 2 this corresponds to a position approximately at the midpoint of the blue line or close to UB4 in the polar configuration (Table 1).

For these calibration data (as for the data later in the analysis), we identify bursts that occurred simultaneously, defined to be within 30 ns, in two or more detectors. The choice of the 30 ns window is due to the time resolution of the detectors. They have a rise time of 13 ns and a decay time of 17 ns. We use a sliding window of 30 ns length to identify the earliest time when we have the largest number of detectors within this time window. If we recorded a saturated signal, we extrapolate missing values with the following procedure: We fit a Gaussian to the signal and take the average of the saturated peak and the Gaussian peak as the estimate of the signal. The Gaussian peak and the saturation value serve as $\pm\sigma$ boundaries for error estimates (black line in Figure 4). It could be argued that a lognormal function should be used, but for reasons already explained (difficulty in determining stop time precisely) we used a Gaussian function.

**Table 1.** Centre Coordinates of Detectors for Each Configuration[a]

| | Azimuth | | | Radial | | | Polar | | |
|---|---|---|---|---|---|---|---|---|---|
| Detector | X | Y | Z | X | Y | Z | X | Y | Z |
| UB1 | 54 | 54 | 70 | 93 | 0 | 62 | 76 | 0 | 94 |
| UB2 | 39 | 67 | 70 | 74 | 0 | 73 | 73 | 0 | 81 |
| UB3 | 20 | 74 | 71 | 59 | 0 | 85 | 70 | 0 | 66 |
| UB4 | 0 | 77 | 72 | 39 | 0 | 91 | 65 | 0 | 53 |
| H1 | 67 | 39 | 70 | 110 | 0 | 51 | 58 | 0 | 39 |

[a]Ground electrode defines origo.

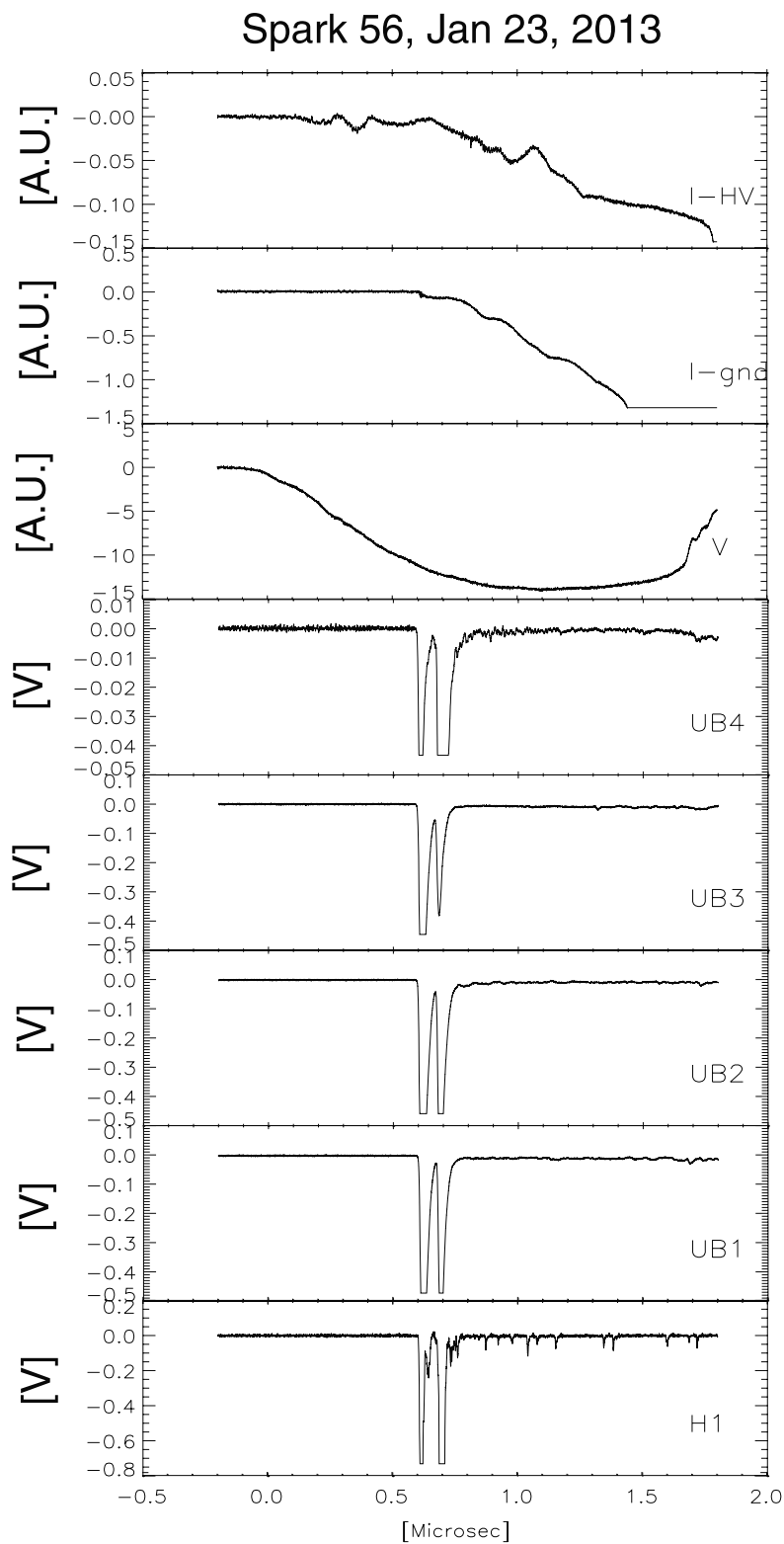

**Figure 3.** One spark where all the detectors have two strong signals and most of them are saturated. Many small peaks are also seen. (first to third panels) The currents at the HV electrode ($I_{HV}$), the ground electrode ($I_{gnd}$), and the Marx generator voltage (V), where −1.5 corresponds to approximately −1 MV. (fourth to eighth panels) The voltage measured by the five detectors. The names of the detectors are identified in the lower right corner of each panel.

In some of the sparks we observe more than one peak in one or more detectors, and in the 75 sparks with signal we now have 154 bursts occurring simultaneously (within 30 ns) in two or more detectors. The results of this intercalibration for the five detectors are shown in Figure 5. The values from the regression line were used for scaling the data using UB1 as reference. For UB4 it can be seen that there are some weak signals that do not follow the regression line. Therefore, values < 0.02 (after calibration) were not used in the analysis.

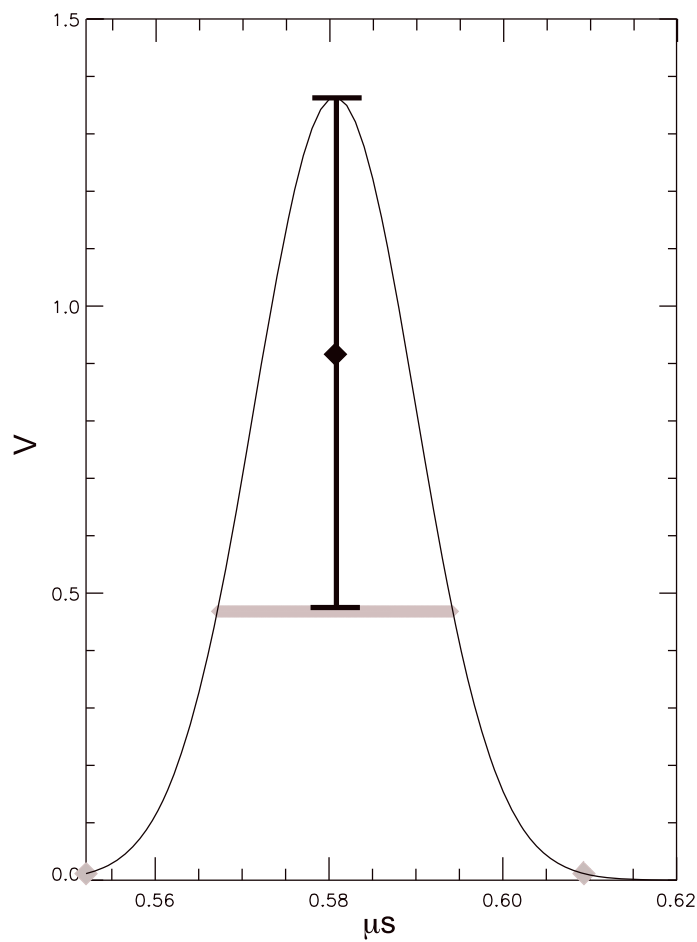

**Figure 4.** Restoration of missing values in saturated signal. Grey diamonds show the start time, the saturated values, and the stop time. Horizontal grey line is the saturation level. Black diamond is the missing value estimate, and black line shows the error bar.

## 5. Analysis

The peak values of the bursts were acquired using the same methods as described in the previous section, namely, (1) simultaneous detection by most detectors within a 30 ns window, (2) restoration of saturated values, (3) scaling all values according to the intercalibration results, using UB1 detector as the reference. For each spark where we have simultaneous peaks in two or more detectors we want to find the most likely source location that would give the measured distribution of relative intensities in the detectors. For a given location we modeled what the expected relative signals should be at the various detector locations given the attenuation of particles in air. The attenuation is quite different for photons and electrons, as described below. We make no assumption about fluence distribution. The relative signal intensities at the detector locations only depend on attenuation and divergence of the particles that give the detected signal.

For this modeling we create a box with the following dimension: $X$:240 cm, $Y$:240 cm, and $Z$:112 cm (see sketch in Figure 6a). This box is divided into 12,600 subboxes (8 cm × 8 cm × 8 cm). From the center position of each of these boxes we model the expected relative signal for each detector for photons of 10 keV and electrons of 200 keV, 300 keV, 400 keV, 500 keV, and 600 keV. The choice of 10 keV photon is due to the following considerations: The attenuation of 3 keV and 15 keV photons varies by a factor of ~50. This means that 10–15 keV photons can reach the detectors from everywhere in the box, while 3–5 keV photons only propagate ~60 cm. A bremsstrahlung spectrum will contain many more 3 keV photons than 15 keV photons but close to the detectors the signal from electrons will be several orders of magnitude larger than the 3–5 keV photons. According to *Cooray et al.* [2009], electrons between 200 keV and 350 keV can be produced by counterstreamers and consequently are expected to give the best match. When comparing estimated relative signals to observation we normalize the two distributions.

### 5.1. Photon Source

For photons of 10 keV we assume an isotropic flux from the source that falls off as $1/r^2$, where $r$ is the distance between source and detector. Attenuation of 10 keV photons is also included but represents a much smaller decrease of the signal strength than the $1/r^2$ effect. Isotropic flux for such low-energy photons should be a reasonable assumption.

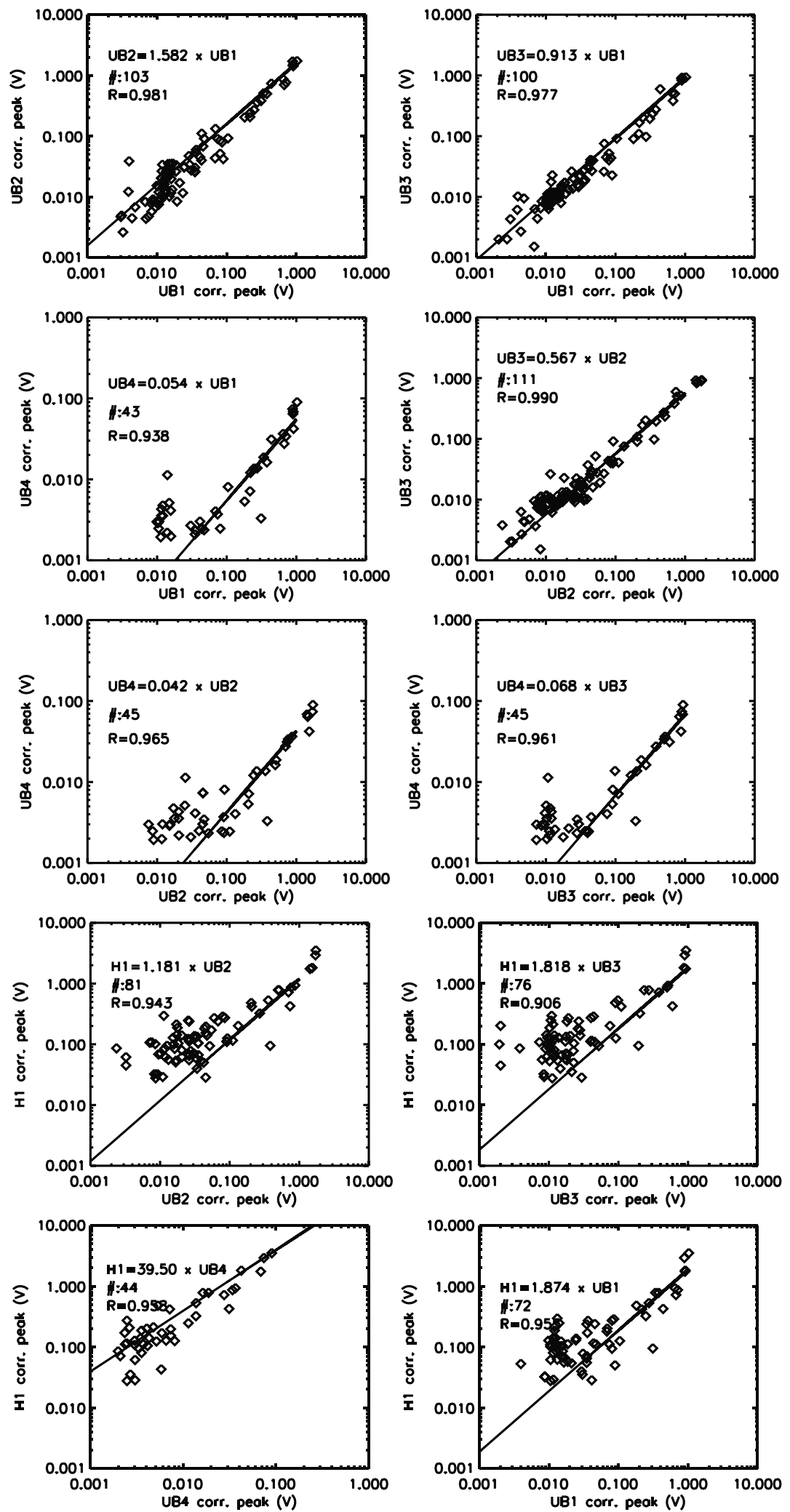

**Figure 5.** Intercalibration based on simultaneous peaks detected from a burst. The saturated peaks are corrected with procedure illustrated in Figure 4.

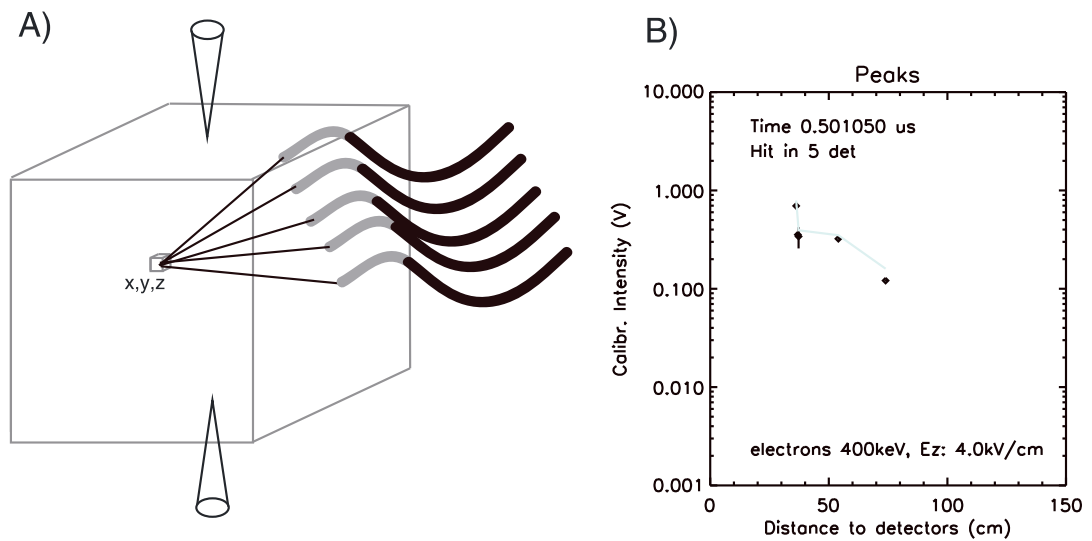

**Figure 6.** (a) Illustration of the modeled signals to the detectors from a source in one subbox element and to the detectors in the radial configuration. Illustration is not to scale. (b) Example of measured and modeled spectra (normalized) from the subbox element using $\chi^2$ statistic to identify the best solution. Diamonds are measured signal, with black vertical line showing the uncertainty from correcting one saturated peak. The grey curve is the modeled best solution.

### 5.2. Electron Source

Modeling of electron propagation (from the source to the detector) is more complex, and we proceed as follows.

1. We assume that the electrons are accelerated in strong localized electric field pointing radially away from the HV electrode. We assume that the extension of this strong localized electric field is smaller than the subbox size of 8 cm we use, in agreement with *Cooray et al.* [2009], who simulated gaps between single streamers from 5 mm to 5 cm.

2. The electric field outside this box was considered uniform in the *z* direction (positive upward), and the decrease/increase of electron energy was estimated depending on whether the electron moves up (decrease) or down (increase) in the electric field to reach the detector. If 200 keV electrons are used in the model, a uniform field of 6 kV/cm is applied, for 300 keV a 5 kV/cm is applied and so on, so the total potential drop will be 800 kV total (for a 107 cm separation of electrodes), which is approximately what the voltage has reached when the bursts are observed.

3. As the collision cross section for electrons in air is energy dependent, a lookup table of attenuation coefficients for every 50 keV electron energy interval was used.

4. As the electrons will undergo many collisions from the source to the detector, they will not move in a straight line but have a random motion. To account for this, we have applied a detour factor of 0.65, which according to *Tabata et al.* [1996] should be valid for 100 keV to 1 MeV. This implies that the electron will have a 1/0.65 longer path length than a straight line.

5. As electrons with energies of several hundreds of keV will be beamed in the direction of the localized electric field pointing radially away from the HV electrode, we need information about the beaming angle, defined as the half-cone angle. To obtain values for this beaming angle a GEANT4 simulation was performed to see the beaming angle at different distances from the source point. The values from this simulation is given in Table 2. Although the beaming angles was estimated for 400 keV electrons, we have used the same beaming angles for electron energies from 200 keV to 600 keV. If the detector is outside the beaming angle, the expected detector signal was set to zero. If the detector is within the beaming angle, a $1/r^2$ divergence factor (in addition to attenuation in air) to the modeled signal is applied, which assumes isotropic distribution within this cone.

Figure 6a shows an illustration of the modeled signal from one subbox to the five detectors in the radial configuration. A set of five signals are estimated from all 12,600 subboxes following the procedure described above for photons and electrons, respectively. All sets of five signals are then compared to the measured signals, and the best fit is determined from $\chi^2$ statistic, as explained in Appendix A1. The goodness of fit will be discussed in section 7. An example of best fit is shown in Figure 6b. This example shows a burst detected when the detectors were in the radial configuration. We modeled the expected signal from 400 keV electrons

| Table 2. Beaming Angle Versus Distance From Source for 400 keV Electrons | |
|---|---|
| Distance (cm) | Half-cone angle (°) |
| 10 | 20 |
| 20 | 30 |
| 30 | 40 |
| 40 | 60 |
| 50 | 50 |
| 60 | 40 |
| 70 | 30 |
| 80 | 20 |
| 90 | 10 |

emitted from two source locations. We will justify the use of two source locations in section 6.2.

## 6. Results

For each of our configurations we performed 100 sparks, and the fractions of sparks where we detected one or more bursts were as follows: azimuthal: 41%, radial: 47%, and polar: 35%. As mentioned in section 3, we expect the source location to be 30 cm to 60 cm from the HV electrode. This is based on the imaging results of counter-streamers from *Kochkin et al.* [2014] and the time coincidence between HV current pulse and X-ray observations [*Kochkin et al.*, 2015].

### 6.1. Photon Source

In Figure 7 we show the model results for sparks performed when the detectors were distributed in an azimuthal configuration, and we assume the source to be 10 keV photons. Of the 100 sparks we had signals in 41, and 35 of these had signals in two or more detectors and could be analyzed. As some sparks had more than one burst, there were 50 bursts that could be analyzed. Based on our best fit analysis between modeled and measured signals (Appendix A1) the best source locations for these bursts have a wide spread and not in the 30–60 cm distance form HV electrode we expected. The average distance from HV electrode is 89.9 cm. The simulated average distances for the two other configurations are listed in Table 4, last row. Since photons of 10 keV are not much attenuated by air, there are no locations within our modeled box that give zero signals in the detectors, which implies that if the fluence of photons is sufficiently high, there should be detectable signals from all sparks. The estimated effects of having a distribution of fluences will be discussed in section 7.1.

### 6.2. Electron Source

By applying the source location determination procedure described in section 5.2, we notice that only a small segment of total volume will give modeled signals in one or more detectors. This is due to beaming and strong attenuation of electrons in air.

In Figures 8a, 8b, 8d, and 8e, this segment between 30 and 60 cm from the HV electrode is indicated by grey diamonds. This is for azimuthal and polar configuration when 300 keV electrons are emitted from each location within 30 cm to 60 cm from the HV electrode. With these configuration and this electron energy, the ratio of this segment (observable volume) to the total volume within 30 cm to 60 cm are 16.7% and 13.6%, respectively. In Figure 8B we have indicated by grey circles the total 30–60 cm volume as seen from above. However, we do not restrict our source determination to be within the 30–60 cm segment, and consequently, we find some source locations both closer than 30 cm and further than 60 cm, but most source locations are found within the 30–60 cm segment. We want to emphasize the following: The 30–60 cm volume is what we expect based on the imaging results of streamers during the rise time of the fourth current pulse at the HV electrode [*Kochkin et al.*, 2014]. The determination of source locations is only due to the limited observable volume when electrons are assumed to be the source. The fact that most of the source locations fall within the expected segment (30–60 cm) is indeed an indication that electrons could be the source population.

However, this also raises a question about how many source locations there could be during one burst (within 30 ns). The ratio of observable segment to total volume (within 30–60 cm) that can give any expected signal is listed for all energies and all configurations in Table 3, and we see that for all configurations, this fraction is less than the fraction of sparks where we detect one or more bursts. Assuming that electrons can be produced anywhere in the total volume between 30 cm and 60 cm from the electrode, this implies that we may have more than one source location in the total volume (grey circle in Figure 8b) during one burst. Just dividing the fraction of sparks with signals (41%, 47%, and 35%, respectively) by the observable volume fractions (from Table 3) we infer that there could be at least two, six, or three source points in the total volume for one burst. Another possibility is that some signals are due to electrons and some are due to photons. This will be discussed.

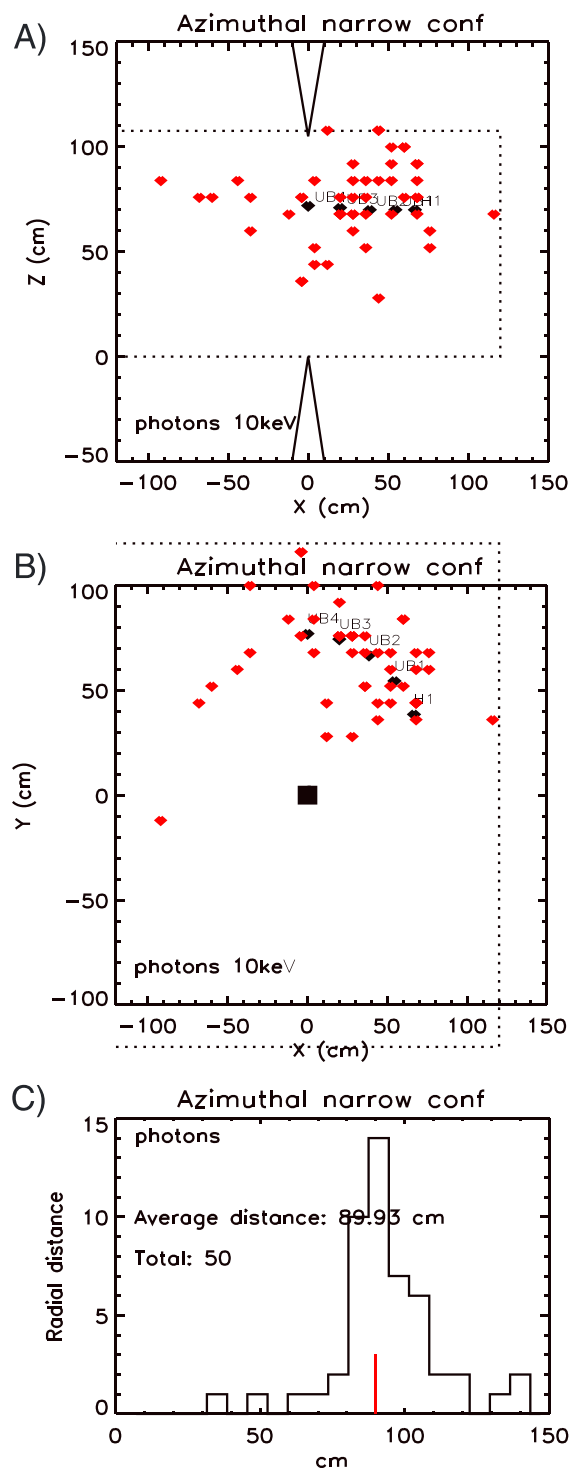

**Figure 7.** Modeled source location for 10 keV photons for the azimuthal configuration. (a) *XZ* view, (b) *XY* view, and (c) distribution of distances from HV electrode to source location.

Furthermore, due to the beaming of electrons, no source point gives modeled signals in all five detectors. However, we often measure signals in all five detectors simultaneously. This means that not only do we have several points of source simultaneously in the total volume but also do we have several within the observable volume (segment) that can give signal in one or more detectors.

Due to this we added a sixth step in our procedure described in subsection 5.2:

1. For all the points predicting one or more signals in the detector we made an additional set of predicted signals from all possible combination of two source points that each predicts one or more signals in the detector. As already emphasized, these source locations are not restricted to the volume indicated by grey diamonds in Figures 8a, 8b, 8d, and 8e but cannot be far from the detectors, as the electrons are effectively absorbed in air.

The most likely source locations are shown with red diamonds in Figures 8a, 8b, 8d, and 8e. As we have not put any restrictions on where the expected source points should be, there are a few that are outside the grey diamond area, but as the radial distributions (Figure 8c and 8f) show, they are fairly close to the expected 30–60 cm range.

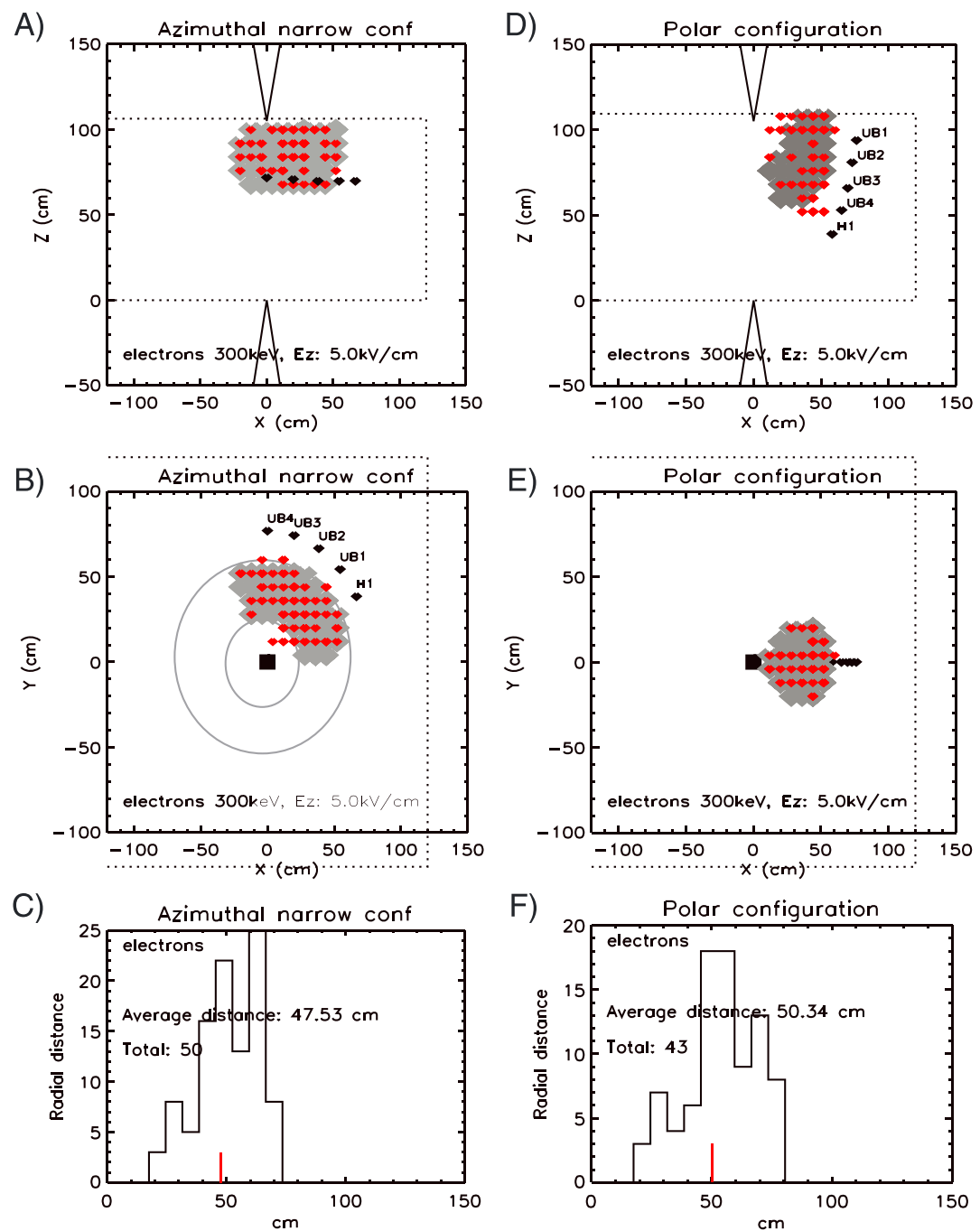

**Figure 8.** Modeled source location for 300 keV electrons. Azimuthal configuration. (a) *XZ* view, (b) *XY* view, and (c) distribution of distances from HV electrode to source location. (d–f) Same as Figures 8a–8c but for polar configuration. In Figure 8b we have indicated the total volume within 30–60 cm with grey ellipses.

## 7. Discussion

We should emphasize that the model we use to predict signals in the detectors is not a detailed Monte Carlo model. However, for predicting most likely locations for a source of isotropic photons of 10 keV, we do believe that the model is quite accurate. This is because the attenuation of photons is quite small for these short distances, and a $1/r^2$ is the most important. For such low-energy photons the assumption of isotropic flux should also be reasonable. For electrons, which are beamed, have energy-dependent attenuation, experience many collisions which leads to a random propagation, and are also affected by the electric field outside the acceleration region, the model is less accurate. However, we have tried to take all these effects into account, and as such, we do believe the model gives us some qualitative indications of what are the most likely source locations, also for electrons. In Appendix A1 we describe how we apply $\chi^2$ statistic to estimate the goodness of our model, as well as the uncertainty of the model source location.

### 7.1. Photon Source

We see from Figure 7 (only showing the azimuthal configuration) that the source locations are more distributed than expected. This is quantified in Table 4 for all three configurations and the average location of source points is ~90 cm away from the electrode for all configurations. This is not quite consistent with the location predicted by the streamer images by *Kochkin et al.* [2014].

Table 3. Fraction (%) of Observable Volume Between 30 cm and 60 cm From the HV That Would Give Signal in One or More Detectors

| Particle Energy (keV) | Configuration | | |
|---|---|---|---|
| | Azimuthal (%) | Radial (%) | Polar (%) |
| el 200 | 10.2 | 3.4 | 12.6 |
| el 300 | 16.7 | 7.2 | 13.6 |
| el 400 | 18.0 | 8.0 | 13.6 |
| el 500 | 18.0 | 8.0 | 13.6 |
| el 600 | 18.0 | 8.0 | 13.6 |
| ph 10 | 100 | 100 | 100 |
| Sparks with signal | 41 | 47 | 35 |

Following the procedure described in Appendix A1 to evaluate how good our model is when 10 keV photons is assumed as source population, we do not obtain acceptable reduced $\chi^2$ values (∼1) for any of the configurations. As seen from Table 4 the median values ranges from 28 to 102, and only a small fraction (∼10%, not listed) of analyzed bursts gives reduced $\chi^2 < 2$ (∼$1 + \sigma_{\chi^2}$). This is one more indication that our measured signals are not produced by photons. However, we should notice that if we had included some limits on the fluence of photons, it would probably have affected our results.

To further test the hypothesis that the source of our detected signals are photons, we have performed a probability analysis based on the results from *Carlson et al.* [2015]. They analyzed data obtained from the same spark experiments that we report, but from two LaBr detectors, each with an effective area of 11.3 cm$^2$, located ∼200 cm from the vertical axis between the electrodes. Due to this large distance and the fact that the detector had a thin aluminum shield it is quite unlikely that these detectors would see electrons. Even 600 keV electrons would be effectively stopped after 130 cm (with a detour factor of 0.65) and shorter distance for lower electron energies. We can therefore assume that these detectors are only measuring photons and due to the aluminum shielding only photons >30 keV can enter the detector.

Using a statistical approach, *Carlson et al.* [2015] found that the fluence behaves as a power law with index −1.29 spanning 3 orders of magnitude and that the average energy of the photons, assuming an exponential energy distribution, is 86 keV. We have used these results as input to estimate the probability of our detectors to measure any photon burst produced by the spark. As our detectors were only sensitive to photons in the energy range 3–15 keV (see Figure 1) we have estimated the probability for detection for both exponential energy distribution (as *Carlson et al.*, 2015, assumed) and for a bremsstrahlung spectrum (∼1/E). Attenuation of photons were also accounted for. The probability analysis is explained in more detail in Appendix A2. The estimated probability to see any photon burst during our experiments is 10% for exponential energy distribution and 16% for bremsstrahlung spectrum. This should be compared to the fraction of sparks with signals (41%, 47%, and 35%) we recorded. Based on this analysis we can not exclude the possibility that 20%–45% of our detected bursts are due to photons, but we can also conclude that the majority of our detected signals are not produced by photons.

Table 4. Average Distance From HV Electrode to Best Source Location[a]

| Particle Energy (keV) | Configuration | | |
|---|---|---|---|
| | Azimuthal (cm; $\chi^2_{med}$) | Radial (cm; $\chi^2_{med}$) | Polar (cm; $\chi^2_{med}$) |
| el 200 | 52.9 (0.75) | 43.9 (160) | 56.9 (3.5) |
| el 300 | **47.5 (0.75)** | 48.6 (56) | **50.3 (0.88)** |
| el 400 | 52.7 (0.75) | 47.4 (55) | 55.3 (2.7) |
| el 500 | 54.1 (0.75) | 46.2 (32) | 55.7 (5.7) |
| el 600 | 52.6 (0.75) | 47.1 (40) | 53.0 (2.9) |
| ph 10 | 89.9 (28) | 91.7 (102) | 88.4 (96) |

[a]Median reduced $\chi^2_{med}$; acceptable solutions are in bold.

For the bursts analyzed in this paper, we can also check how many of these bursts were seen by both LaBr detectors and our fiber detectors. Of the total number of detected bursts about half of them where seen by both LaBr and fiber detectors (radial: 48%, polar: 50%, and azimuthal: 51%). As our fiber detectors are less sensitive to photons than LaBr and could have measured electrons while LaBr measured photons, this is consistent with the probability analysis predicting that 20%–45% of our detected bursts could be produced by photons. This is another indication that the majority of our detected signals are not from photons.

Both the distributed source location, the average modeled distance from the HV, the probability estimates, and that only half of the detected bursts where detected by both LaBr, and our fiber detectors indicate that the majority of detected signals by the fiber detectors are not produced by photons.

### 7.2. Electron Source

By following the procedure outlined in Appendix A1, we find that for two of the configurations, azimuthal and polar, we obtain acceptable reduced $\chi^2$ values (~1). For the 50 analyzed bursts with the detectors in azimuthal configurations the median and mean value of reduced $\chi^2$ are 0.75 and 1.16, respectively, when 300 keV electrons are assumed as the source population. Of these 50 compared measured and modeled signals, 42 have reduced $\chi^2 < 2$ (~$1 + \sigma_{\chi^2}$). For 34 of these comparisons there are less than two surrounding subboxes that gives reduced $\chi^2 < 2$. This indicates that the uncertainty of the source location returned by the model is close to the size of the subbox we use, namely 8 cm. The other electron energies also give similar median values of reduced $\chi^2$ (see Table 4) but with slightly larger mean values.

For the polar configuration we also obtain median reduced $\chi^2$ values close to 1 (0.88) when 300 keV electrons are assumed as source population, although some outliers gives a large mean value. Of 43 bursts analyzed, 26 give reduced $\chi^2 < 2$. For 19 of the bursts, there are less than two surrounding subboxes that gives reduced $\chi^2 < 2$, indicating that the uncertainty for the modeled source location is about 8 cm for the vast majority of the comparisons. All the other electron energies give reduced $\chi^2$ values with large median and mean values.

For the radial configuration we do not obtain any acceptable reduced $\chi^2$ values (for 70 bursts compared), the lowest median value is 32 (500 keV electrons) which indicates that the model does not work for this configuration. There are several factors that can explain this. First, we notice that the $\chi^2$ is very sensitive to the estimate of $\sigma_{fit}$. Second, our simplistic model does not consider any fluence distribution, energy distribution, or angular distribution of electrons at the source, and our electric field geometry is also simplified. For the latter we assume that the electric field that accelerate the electrons is pointing radially away from the electrode. For the volume we consider (30 cm–60 cm) this should be a reasonable assumption. However, we have not taken into account that the path of the electrons would also be affected slightly by the background field, but the uncertainty introduced by this simplification should be less than the size of the boxes we use. We assume a very large number of monoenergy electrons and one initial direction (radially away from the electrode), and if the electrons can propagate from source location to detector, they will be detected. In reality the electrons may have an initial angular distribution when accelerated in the local electric field of counter-streamers and also limited fluence. Even a sophisticated Monte Carlo would have to make assumptions about these unknowns. Nevertheless, our simplistic model give us some support for 300 keV electrons being a likely source population.

In section 6.2 we explained why we had to use two source locations to get signals in all five detectors for some of the events. We justified this by the fraction of sparks with signals (41%, 47%, and 35%) are larger than the volume fraction (due to beaming and attenuation) that could produce signals in at least one detector and suggested that in case of electrons, there need to be two, six, or three source locations in the total volume (30–60 cm from the HV electrode) within 30 ns. This further suggests that for some of the sparks (to get signal in all detectors), more than one source location is needed within the observable volume fraction listed in Table 3.

From Table 4 we see that the average location of source points, for all the electron energies, are within the expected radial distance from the HV electrode (30–60 cm). This is the strongest indication from this simulation that most of the signals are indeed from electrons. Given these close source locations, we cannot rule out that there could also be some photons hitting the detector, but as already pointed out the intensity of photons (from bremsstrahlung production) should be several orders of magnitude lower than from the electrons coming from this close distance.

We have run the simulations for electron energies from 200 keV to 600 keV, and they all give reasonable results regarding source locations, but the reduced $\chi^2$ values indicate that we only have a reasonable model when 300 keV electrons are assumed as the source population. We do see that the volume fractions for 200 keV (Table 3) are smaller than the other energies. This is because the electrons of this energy must be produced closer to the detector than for higher energies. Electrons of lower energies (≤100 keV) will not be detected by our detectors (see Figure 1). The highest energy we have simulated is 600 keV. Given that we only have a 800 kV potential total available, we believe it is not realistic to produce electrons with more than 600 keV, and even that is probably unrealistic high.

Based on this discussion we will conclude that the majority of signals we observe are produced by electrons and most likely with energies of ∼300 keV. This is consistent with the predictions from *Cooray et al.* [2009] that electrons with energies up to 360 keV can be produced by counterstreamers. The electrons have to be produced close to the detectors. They will also produce photons, but the total energy deposited by these photons would be several orders of magnitude smaller than from the electrons. However, our analysis also indicates that 20%–45% of our signals could be produced by photons coming from distances further away from the detectors.

The X-rays produced by 300 keV electrons would have an average energy of 70 keV. This is smaller than some of the energies reported earlier, 230 keV [*Dwyer et al.*, 2008] and 200 keV [*Kochkin et al.*, 2015], but close to the 86 keV reported by [*Carlson et al.*, 2015] and consistent with the 30 keV–150 keV reported by *Dwyer et al.* [2005].

## 8. Summary

We have analyzed 300 sparks performed in the laboratory, where five very thin scintillating plastic fiber detectors were arranged in three different configurations. The detectors were designed to be more sensitive to ≥200 keV electrons than photons. For each spark we have identified simultaneous signals in one or more detectors and corrected for saturated signals. We have then modeled the most likely source location assuming 10 keV photons and electrons of 200 keV to 600 keV. We have applied $\chi^2$ statistic to estimate the goodness of our model as well as the uncertainty of modeled source location. We have performed a probability analysis based on results from *Carlson et al.* [2015] to estimate the probability of our detectors to detect photons in the energy range our detectors were sensitive to. This estimate have further been compared to the fraction of detected sparks with bursts that where seen by both LaBr and our detectors.

Our conclusions regarding the three questions we wanted to address (see end of section 3 ) are the following:

1. The majority of the signals we have measured is produced by relativistic electrons not far away from the detectors. We can not rule out that a significant fraction (20%–45%) could be produced by low-energy photons from further away.
2. The simulated source locations of the electrons are consistent with the region of counterstreamers 30 cm to 60 cm from the HV electrode.
3. Our model indicates that the electron energies we detect are most likely ∼300 keV.

To our knowledge, this is the first study that reports direct measurements of relativistic electrons from sparks in the laboratory.

## Appendix A

### A1. The Goodness of the Model and Uncertainties of Source Location

To test the goodness of our model we calculate the test statistic for all subboxes (source points) with distance $r_i$ from each detector for each observed burst.

$$\chi^2(r) = \frac{1}{D_f} \sum_i \frac{(Y_{\text{obs}}(i) - Y_{\text{fit}}(i, r_i))^2}{\sigma_{\text{tot}}(i)^2} \quad \text{(A1)}$$

where $D_f$ is the number of degrees of freedom, $Y_{\text{obs}}(i)$ is the observed signal in each detector, $Y_{\text{fit}}(i, r_i)$ is the modeled signal at each detector from distance $r_i$, and $\sigma_{\text{tot}}(i)$ is the total error:

$$\sigma_{\text{tot}}(i)^2 = \sigma_{\text{obs}}(i)^2 + \sigma_{\text{fit}}(i, r_i)^2 \quad \text{(A2)}$$

The $\sigma_{\text{obs}}(i)$ is given by the noise in our measured signal and the error resulting from the correction for saturated peaks. The $\sigma_{\text{fit}}(i)$ is estimated in the following way: First, we estimate the reduced $\chi^2$ (A1) by using only $\sigma_{obs}(i)$ as $\sigma_{tot}(i)$. The subbox that returns the minimum reduced $\chi^2$ value is chosen as the most likely source point, $r_{0,i}$. Then we estimate the $\sigma_{fit}(i, r_{0,i})$ for this source point ($r_{0,i}$) by using the $Y_{fit}(i, r_{n,i})$ from the number of nearest-neighbor subboxes ($n_{sur}$) surrounding this point.

$$\sigma_{fit}(i, r_{0,i})^2 = \sum_{n}^{n_{sur}} \frac{(Y_{fit}(r_{n,i}, i) - Y_{fit}(r_{0,i}, i))^2}{n_{sur}} \quad (A3)$$

For one source point $n_{sur}$ equals 6, while for two source points it is 12. The reduced $\chi^2(r_0)$ is then recalculated by including $\sigma_{fit}(i, r_{0,i})^2$ in $\sigma_{tot}(i)^2$ (A1). For nonsaturated peaks the $\sigma_{obs}$ is smaller than $\sigma_{fit}$, by a factor 0.1–1, while peaks corrected for saturation $\sigma_{obs}$ is usually larger by a factor 1–10. The reduced $\chi^2$ value should be $\sim 1$ and $\chi^2 < 2$ is $\sim 1 + \sigma_{\chi^2}$.

### A2. Photon Detection Probability

The purpose of this appendix is to estimate how many bursts would be observed in the current experiment if they were made up exclusively out of photons. This is done by using the results of *Carlson et al.* [2015] and applying them to the current experiment. They analyzed data obtained from the same spark experiments that we report, but from two LaBr detectors, each with an effective area of 11.3 cm$^2$, located $\sim$200 cm from the vertical axis between the electrodes. Due to this large distance and the fact that the detector had a thin aluminum shield it is quite unlikely that these detectors would see electrons. In that experiment, the individual bursts were not distinguished, but the photons were summed over all bursts in a spark. We will denote by $n$ the total number of photons in a detector produced in a spark.

*Carlson et al.* [2015] measured the power law distribution for $n$:

$$p_\lambda(n) = \frac{(-\lambda - 1)n^\lambda}{n_{min}^{\lambda+1} - n_{max}^{\lambda+1}}, \quad n_{min} < n < n_{max}$$

with parameters $n_{min} = 0.022$, $n_{max} = 110$, and $\lambda = -1.29$. The fraction of sparks with bursts (one or more photons detected; the number of photons is Poisson distributed) is

$$f = 1 - \int_{n_{min}}^{n_{max}} e^{-n} p_\lambda(n) \, dn \approx 1 - \frac{n_{min}^{\lambda+1} e^{-n_{min}} - \Gamma(\lambda + 2, n_{min})}{n_{min}^{\lambda+1} - n_{max}^{\lambda+1}} \quad (A4)$$

where $\Gamma(s, x)$ is the upper incomplete gamma function, and we neglected $\Gamma(\lambda + 1, n_{max})$.

For a combination of two detectors in *Carlson et al.* [2015] experiment we must rescale $n_{min,max} \to 2n_{min,max}$. We may check that the expected fraction of sparks with a signal is then $f = 45\%$, which is close to 43% reported by *Carlson et al.* [2015].

In the presented experiment, the number of photons entering the detector located at distance $r$ from the source is

$$n = \frac{1}{4\pi r^2} \int F_A A(E) e^{-\mu_{air}(E) r} I(E) \, dE \quad (A5)$$

which includes the photon energy-dependent area $A(E)$ plotted in Figure 1, attenuation in air $\mu_{air}(E)$ [*Hubbell and Seltzer*, 2004] and the photon spectrum $I(E)$ (number of photons produced in a spark per unit energy interval). The factor of $F_A = 2.5$ is obtained from the following considerations: the burst was registered by any single detector out of five, so the detector areas are summed; the fact that for some source locations the detector area is not fully visible may be taken by an extra factor of 0.5. As we see from Figure 1, the appropriate integration limits could be taken approximately from 3 to 15 keV.

In order to estimate the expected fraction of sparks with bursts in the present experiment, let us introduce the ratio for two experiments $R = n_1/n_2$, where $n_2$ is for *Carlson et al.* [2015] experiment and $n_1$ is for the present experiment. Then we may use formula (A4), into which we substitute rescaled values $n_{min,max} \to R n_{min,max}$, to find the expected fraction of sparks with bursts.

We use $r_1 = 100$ cm for the distance from the source to the detector described in the present work. The value of $n_2$ for *Carlson et al.* [2015] experiment is also calculated using equation (A5), with $A(E) \approx 11.3$ cm$^2$ in the limits 30–400 keV and distance $r_2 = 200$ cm, for a single detector ($F_A = 1$), and we may neglect attenuation at these energies. The photon bursts are the same in the two experiments, so $I(E)$ is the same. Since we take the ratio, we do not need to know the absolute value of $I(E)$, just the shape. We consider two different shapes of spectra of photon energies: the bremsstrahlung spectrum presented above and an exponential spectrum of *Carlson et al.* [2015].

1. Bremsstrahlung spectrum, $I(E) \propto E^{-1}$: number of photons in the detector ratio is $R = 0.12$, and the expected fraction of sparks with bursts $f = 16\%$.
2. Exponential spectrum, $I(E) \propto e^{-E/\mu}/\mu$, with $\mu = 86$ keV [*Carlson et al.*, 2015]: $R = 0.05$ and $f = 10\%$.

To account for possible variations in the position of the source in *Carlson et al.* [2015] experiment, we also give results for $r_2 = 250$ cm: $R = 0.19$ and $f = 19\%$ for bremsstrahlung spectrum and $R = 0.08$ and $f = 13\%$ for exponential spectrum, respectively.

**Acknowledgments**
The data described in this paper are available from the authors on request (nikolai.ostgaard@uib.no). This study was supported by the European Research Council under the European Unions Seventh Framework Programme (FP7/2007–2013)/ERCgrant agreement 320839 and the Research Council of Norway under contracts 208028/F50, 216872/F50, and 223252/F50 (CoE).